\documentclass[sigconf,natbib=true]{acmart}
\usepackage{graphicx} 
\usepackage{caption}
\usepackage{subfigure}
\usepackage{multirow}
\usepackage{url}
\usepackage{balance}

\AtBeginDocument{%
  \providecommand\BibTeX{{%
    \normalfont B\kern-0.5em{\scshape i\kern-0.25em b}\kern-0.8em\TeX}}}

\copyrightyear{2023}
\acmYear{2023}
\setcopyright{acmcopyright}\acmConference[WSDM '23]{Proceedings of the
Sixteenth ACM International Conference on Web Search and Data
Mining}{February 27-March 3, 2023}{Singapore, Singapore}
\acmBooktitle{Proceedings of the Sixteenth ACM International Conference on
Web Search and Data Mining (WSDM '23), February 27-March 3, 2023,
Singapore, Singapore}
\acmPrice{15.00}
\acmDOI{10.1145/3539597.3570380}
\acmISBN{978-1-4503-9407-9/23/02}

\begin{document}

\title{Slate-Aware Ranking for Recommendation}

\author{Yi Ren}
\authornote{Both authors contributed equally to this research.}
\authornote{Corresponding author.}
\affiliation{%
  \institution{Tencent }
  \city{Beijing}
  \country{China}}  
\email{henrybjren@tencent.com}

\author{Xiao Han}
\authornotemark[1]
\affiliation{%
  \institution{Tencent}
  \city{Beijing}
  \country{China}}  
\email{alisahan@tencent.com}

\author{Xu Zhao}
\affiliation{%
  \institution{Tencent }
  \city{Beijing}
  \country{China}}    
\email{xuzzzhao@tencent.com}

\author{Shenzheng Zhang}
\affiliation{%
  \institution{Tencent}
  \city{Beijing}
  \country{China}}  
\email{qjzcyzhang@tencent.com}

\author{Yan Zhang}
\affiliation{%
  \institution{Tencent}
  \city{Beijing}
  \country{China}}  
\email{zyzn5288@126.com}

\renewcommand{\shortauthors}{Yi Ren, Xiao Han, Xu Zhao, Shenzheng Zhang, \& Yan Zhang}

\begin{abstract}
We see widespread adoption of slate recommender systems, where an ordered item list is fed to the user based on the user interests and items' content. For each recommendation, the user can select one or several items from the list for further interaction. In this setting, the significant impact on user behaviors from the mutual influence among the items is well understood \cite{bello2018seq2slate,pang2020setrank,deng2018ad,zhuang2018globally,pei2019personalized,wang2019sequential,jiang2018beyond,zhao2017deep,gong2019exact,wei2020generator,feng2021grn,ai2018learning,ai2019learning,liu2021variation,feng2021revisit}. The existing methods add another step of slate re-ranking after the ranking stage of recommender systems, which considers the mutual influence among recommended items to re-rank and generate the recommendation results so as to maximize the expected overall utility. However, to model the complex interaction of multiple recommended items, the re-ranking stage usually can just handle dozens of candidates because of the constraint of limited hardware resource and system latency. Therefore, the ranking stage is still essential for most applications to provide high-quality candidate set for the re-ranking stage. In this paper, we propose a solution named Slate-Aware ranking (\textbf{\textit{SAR}}) for the ranking stage. By implicitly considering the relations among the slate items, it significantly enhances the quality of the re-ranking stage's candidate set and boosts the relevance and diversity of the overall recommender systems. Both experiments with the public datasets\footnote{Code link is: https://github.com/BestActionNow/Slate\_Aware\_Ranking} and internal online A/B testing are conducted to verify its effectiveness.   

\end{abstract}


\maketitle

\vspace{-0.2cm}
\section{Introduction}
\begin{figure}
\setlength{\abovecaptionskip}{0.0cm} 
	\centering\textbf{}
	\includegraphics[scale=0.13]{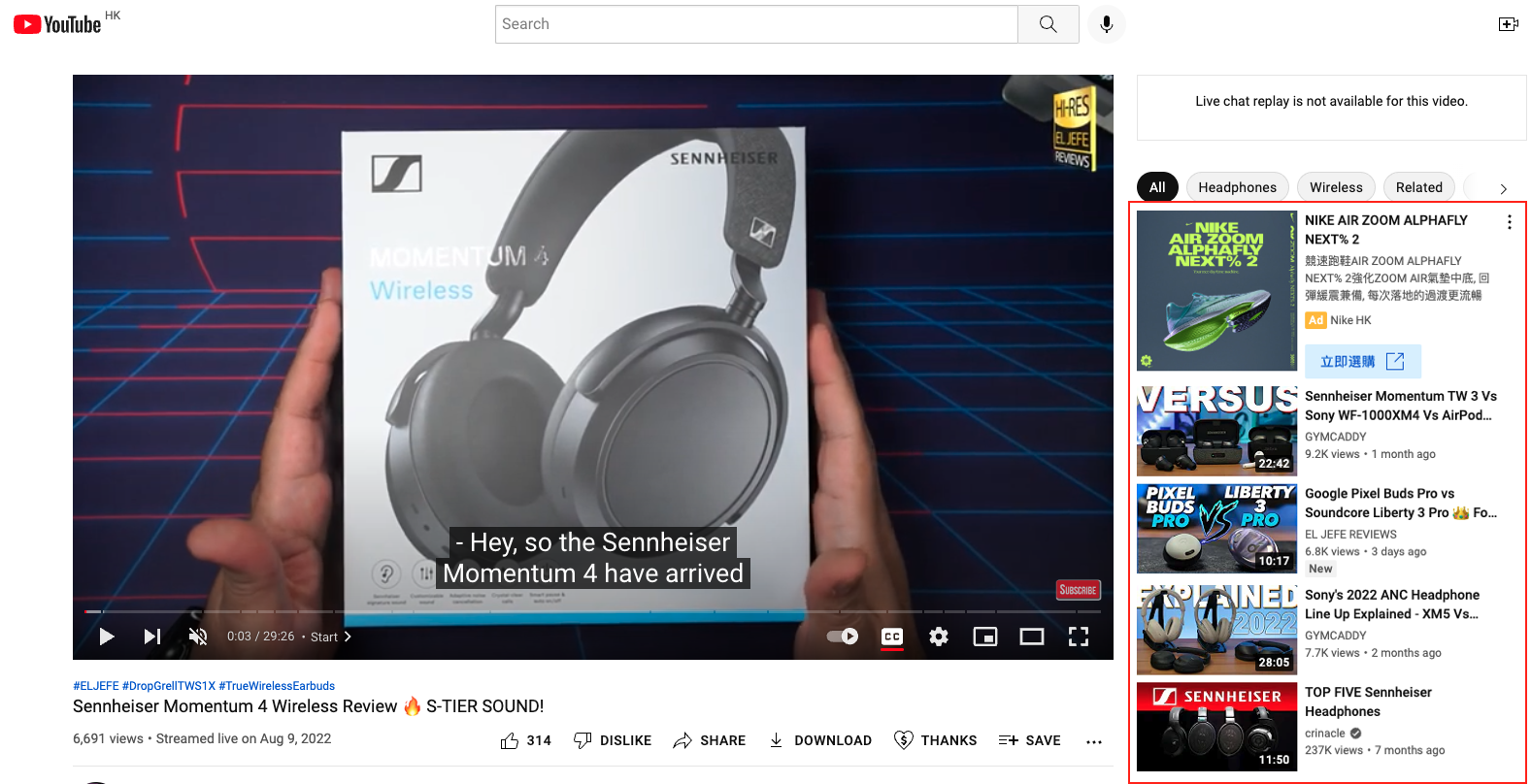}
	\caption{Watch Next Recommendation on Youtube}
	\label{fig:youtube_rec}
\end{figure}

In this era of information explosion, recommender systems are useful tools to filter out valuable information for the users and already impact almost every facet of people’s lives. Typically, for each user request, recommender systems select a slate of items for presentation from the large candidate set based on its understanding on the user's intention. There are two main reasons for returning a slate of items rather than a single item. First, in many recommendation scenarios, the user can observe multiple items in the screen at once. Moreover, the overload of the recommender systems is greatly alleviated by lowering the QPS (Query-Per-Second) of interactions between the users and system servers. After receiving the recommendation slate, the user can scroll the screen and consume the items relevant to his current interest. A new user request will be issued once the slate of items are all presented. Please refer to the "watch next recommendation" scenario on Youtube displayed at the right pane of Figure \ref{fig:youtube_rec} for a representative slate recommendation scenario.     

Generally speaking, typical recommender systems take the multi-stage design and consist of at least two steps \cite{covington2016deep}, namely matching and ranking. For some later systems \cite{wang2020cold}, additional steps, including matching, pre-ranking, ranking and re-ranking, are introduced for better trade-off between recommendation performance and system latency.  From matching to re-ranking, more and more powerful models and comprehensive features are utilized yet with a decreasing candidate size. The steps of matching and pre-ranking usually utilize highly efficient methods to filter out relevant items from a vast candidate pool and care coverage more than precision. What's more, with limited hardware resource and system latency, the re-ranking stage normally just handles dozens of candidates by design. Hence, the ranking stage, which selects top tens of items from a pool of hundreds to thousands with high precision, plays an essential role for the recommendation performance.  

Traditionally, the models in the ranking stage compute a uni-variate score for each given user-item pair and construct the slate based on the order of item scores \cite{cremonesi2010performance}. As the top ranked items take more chance to
be observed or preferred, the overall efficiency should be optimized in this way if we assume the user's preference for a item is stable. Nevertheless, it is well known that the user interaction and feedback depends on not only the corresponding item's quality and relevance but also the contextual items' impact \cite{bello2018seq2slate,pang2020setrank,deng2018ad,zhuang2018globally,pei2019personalized,wang2019sequential,jiang2018beyond,zhao2017deep,gong2019exact,wei2020generator,feng2021grn,ai2018learning,ai2019learning,liu2021variation,feng2021revisit}. As a result, the traditional models in the ranking stage, which just accept a given user-item pair's features to learn a uni-variate scoring function, is incapable of taking into account the complex influence exerted by the surrounding items. And the recommendation performance tends to be sub-optimal. 

To tackle this challenge, various algorithms have been devised and implemented in the re-ranking stage \cite{bello2018seq2slate,pang2020setrank,zhuang2018globally,pei2019personalized,wang2019sequential,zhao2017deep,gong2019exact,wei2020generator,feng2021grn,ai2018learning,ai2019learning,feng2021revisit}. Acting as the last step after the matching and ranking stage, the re-ranking stage obtains the top ranking items from the ranking stage as candidates and refines into the final recommendation results. Admittedly, these re-ranking methods achieved success in many industrial applications. Nevertheless, to account for the complex relations among the displayed items, these models are usually much more complex than the models in the ranking stage. Considering the limited system latency, the re-ranking models usually can only handle dozens of items ranked top by the ranking stage. Thus, it is essential for the ranking stage to provide highly relevant yet diverse top item set. Nonetheless, as the ranking stage is oblivious to the overall recommendation list and does not consider the cross item impact, the re-ranking stage are likely to receive unsatisfactory candidate set, thereby negatively affecting the overall performance of recommender systems.  

There are some challenges for addressing the issue of cross item impact in the ranking stage. In contrast to the re-ranking stage, the ranking stage just provides competitive candidates to the next stage and cannot determine the final recommendation results. Therefore, we cannot use the slate-wise features in the ranking stage for model training. Moreover, with much larger candidate set, it is also infeasible to deploy these sophisticated and time-consuming models used in the re-ranking stage, such as the sequential generation algorithms \cite{bello2018seq2slate,zhuang2018globally,gong2019exact}, to the ranking stage.

To model the mutual influence among the items of a recommendation slate in the ranking stage, one viable solution is to borrow the ideas from the works of privileged features, which are defined as the features only available for training, and distillation \cite{lopez2015unifying,xu2020privileged}. Besides the base ranking model, a teacher model utilizing both the base ranking features and the slate-wise features can be trained. With knowledge distillation, valuable information can be transferred from the teacher model to the base ranking model served online. But this approach need roughly double the training resources. In addition, without specialized attention to the privileged features, the distillation methods may generate sub-optimal performance, which is verified by our experiment. 

We propose the method of Slate-Aware Ranking (\textbf{\textit{SAR}}), which utilizes the whole recommendation slate's information during training, for the ranking stage. Specifically, with an encoder network, we construct the slate-wise features based on the whole item sequence in a slate and encode them into the latent space, which is then processed by a decoder network and concatenated to the input of upper ranking modules so as to help to minimize the training loss. Meanwhile, with another encoder network, we also encode the user features to an embedding vector that matches the slate's embedding well. The rationality is that the recommender systems are inclined to recommend results with homogeneous patterns based on the user's previous behaviors, which can be memorized by the user encoder network. As we cannot figure out the actual recommendation slate during prediction, we use the embedding vector from user features instead. The most similar works come from \cite{liu2021variation,jiang2018beyond}, which also encode the information of the whole slate to the latent space. But as a generation model, they can just use the vector-product based DNN model with limited expressive capacity. Previous works \cite{wang2020cold,zhu2018learning,xu2020privileged} have shown the obviously superior performance of incorporating complex deep models over vector-product form networks. Moreover, the generative methods show stochastic behavior because of the Reconstruction-Concentration dilemma \cite{liu2021variation}. Finally, they need input the best reward, which is hard to define for multi-task applications, as a generation condition during inference. 

We summarize our main contributions below.
\vspace{-0.1cm}
\begin{itemize}
\item We design an effective solution to implicitly model the mutual influence on user behaviors among the slate items in the ranking stage, which generates significantly better recommendation relevance and diversity by boosting the quality of the re-ranking stage's candidate set. 
\item Compared with the distillation and privileged feature methods \cite{lopez2015unifying,xu2020privileged}, we achieve significant performance gains with around half training resources. 
\item We conduct extensive offline and online experiments to evaluate and understand the effectiveness of our method. Online A/B testing in one of the world’s largest content recommendation platforms shows significant improvement of the business metrics.
\end{itemize}

\begin{figure*}[!htbp]
\setlength{\abovecaptionskip}{-0.0cm} 
	\centering\textbf{}
	\includegraphics[scale=0.39]{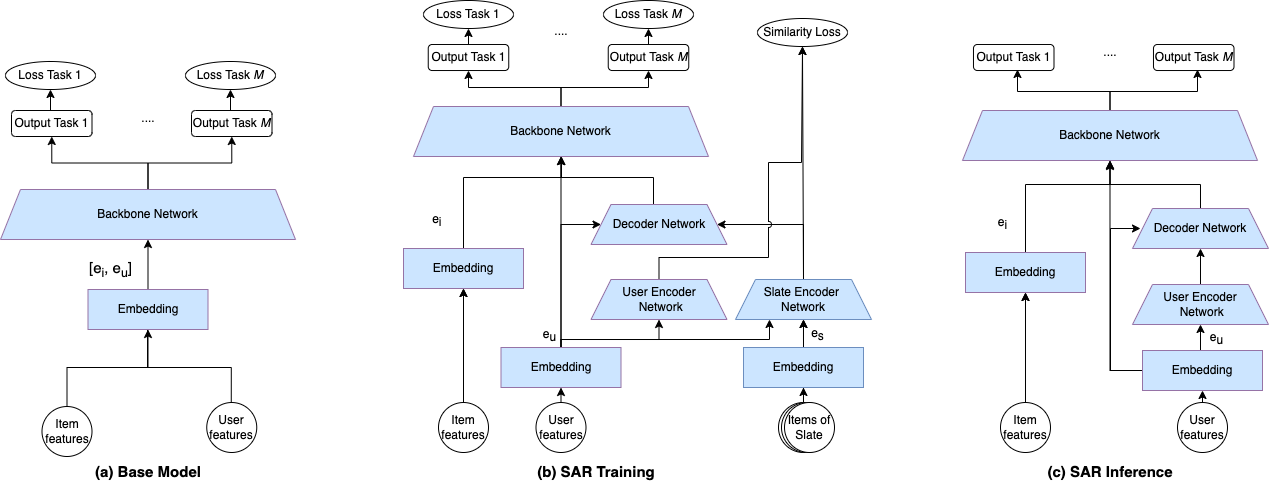}
	\caption{Model Architecture of SAR}
	\label{fig:1}
\end{figure*}

\vspace{-0.2cm}
\section{Related Work}
\subsection{Re-ranking Models for Slate Recommendation}
To model the mutual influence among slate items, various algorithms have been devised and implemented in the re-ranking stage. Acting as the last step after the matching and ranking stage, the re-ranking stage obtains the top ranking items from the ranking stage as candidates and refines into the final recommendation results.  

These re-ranking methods can be roughly classified into three categories. The first category comprises the methods of \textbf{\textit{one-step generator}}, which extract both item-wise features and slate-wise features for each item, predict its relevance score and perform greedy ranking. Though DLCM \cite{ai2018learning}, PRM \cite{pei2019personalized}, GSF \cite{ai2019learning} and SetRank \cite{pang2020setrank} leverage different modelling methods, they all belong to this type. Admittedly, for some methods \cite{ai2018learning}, the scoring function may vary based on the permutation of the initial ranking list, all of them \cite{ai2018learning,pei2019personalized,ai2019learning,pang2020setrank} are not sensitive to the order of the final recommendation slate. Second, for the methods of \textbf{\textit{sequential generator}} \cite{bello2018seq2slate,zhuang2018globally,gong2019exact}, besides the slate-wise feature extraction and encoding, these methods explicitly model the sequential impact on the current item from the previously chosen ones with uni-directional sequential models, such as LSTM networks \cite{greff2016lstm}. Nonetheless, as a sequential generation model, they have to ignore the current item's influence on the previous ones, which does not adhere to the true distributions. Finally, for the methods of \textbf{\textit{generator and evaluator}} \cite{wang2019sequential,feng2021revisit,wei2020generator,feng2021grn}, they initially generate multiple slate candidates with different methods mentioned above and use a separate evaluator model to compute the expected utility of each slate. As the evaluator is able to utilize the bi-directional models, such as bi-directional LSTM and transformer networks \cite{vaswani2017attention}, they are able to capture both the forward and backward impact among the recommended items so as to ensure better recommendation quality.  

Please note that there are explicit reasons why we cannot directly apply these algorithms to the ranking stage. First, unlike the re-ranking stage, the ranking stage cannot decide the final recommendation slate. As a result, all of the algorithms with sequential generation function \cite{wang2019sequential,feng2021revisit,wei2020generator,feng2021grn,bello2018seq2slate,zhuang2018globally,gong2019exact}, which select the current item based on the previously chosen items, cannot be used at all. Furthermore, many algorithms mentioned above, such as \cite{ai2018learning,pei2019personalized}, rely on the item ranking from the previous stage, which is often not available for the ranking stage. Finally, these algorithms are too complex and time-consuming to be practically applied in the ranking stage with much bigger candidate set.
\vspace{-0.2cm}
\subsection{Privileged Features and Distillation}
For recommendation performance, it is essential for us to ensure the consistency between offline training and online inference. However, with this constraint, many discriminative features, which are available at training but not for inference, must be ignored for model training. In \cite{lopez2015unifying,xu2020privileged}, they define the features only available for training as privileged features. Two models, a student and a teacher model, are trained. Unlike normal model distillation methods \cite{bucilu2006model,hinton2015distilling}, which input the same feature set to the student and teacher models, the teacher model here utilizes all features while the student model here does not depend on the privileged features. During training, besides the original training labels, knowledge distilled from the teacher model is also used to supervise the training of the student model. After model convergence, only the student model is deployed online to serve the user request. As the student model does not rely on privileged features, the consistency between online and offline is guaranteed.    

Taken the slate-wise information as privileged features, we can use PFD \cite{xu2020privileged} to indirectly model the mutual influence among the items of a recommendation slate. Compared with SAR, it roughly doubles the training resources. What's more, SAR can utilize the slate-wise information more effectively and achieve better performance, which is verified in our experiments. 
\vspace{-0.2cm}
\subsection{Multi-Task Learning for Recommender Systems}
There are a variety of user interactions in typical recommendation scenario, such as clicking, watch time, liking and commenting. Moreover, single interaction itself are unable to clearly reflect  the user satisfaction. Therefore, the recommender system needs to model multiple user behaviors and further combines there predictions to calculate the final utility score. Under such circumstances, we see wide application of multi-task learning techniques as they can solve multiple related tasks simultaneously and improve learning efficiency and prediction accuracy over the single task methods. 

We briefly introduce related papers here. Hard parameter sharing is the most intuitive MTL structure. For instance, the ESSM model \cite{ma2018entire} shares embedding parameters between the tasks of CTR (Click-Through Rate) and CVR (Conversion Rate) for improving the prediction performance of the sparse CVR task. To alleviate the task conflict and negative transfer issue, Zhao et al. \cite{zhao2019recommending} extend the Multi-gate Mixture-of-Experts model (MMoE) \cite{ma2018modeling} and apply it to learn multiple ranking objectives in Youtube video recommender systems. PLE \cite{tang2020progressive} achieves superior performance for news recommendation by assigning both shared parameters among tasks and task specific parameters. 

In section 3, we will describe how to seamlessly integrate SAR with these multi-task learning methods.

\vspace{-0.2cm}
\section{Methodology}

\begin{table}
\setlength{\belowcaptionskip}{0.1cm}
\begin{tabular}{ll}
\hline
Notation & Description \\
\hline
$u$ & User request with user profile and context \\
$i$ & An recommendation item  \\
$C$ & The candidate set \\
$N$ & The size of the candidate set \\
$M$ & The number of prediction tasks \\
$S$ & The recommended slate chosen from $C$ \\
$K$ & The size of the recommended slate \\
$i_k$ & The $k$th item of $S$ \\
$R(S,u)$ & The reward for recommending $S$ given $u$ \\
$r(i,u)$ & The reward for recommending the item $i$ given $u$ \\
$o_k$ & Observation probability of the $k$th item in the slate \\
$e_u$ & Concatenated embeddings for the user features \\
$e_i$ & Concatenated embeddings for the item features \\
$e_s$ & Concatenated embeddings for the slate-wise features \\
$l_u$ & Encoding results of the user encoder network \\
$l_s$ & Encoding results of the slate encoder network \\
$\lambda$ & The weight for the similarity loss \\
$\omega_m$ & The weight for the loss of the $m$th task \\

\hline
\end{tabular}
\caption{Notations and Descriptions}
\end{table}

In this section, we introduce the \textbf{\textit{SAR}} method in detail. First, we formally introduce the problem definition for slate recommendation. Then, we explain the general design of SAR for the ranking stage. Finally, we elaborate on how to enhance the ranking models with the proposed method. 
\vspace{-0.2cm}
\subsection{Slate Recommendation}
Given a set of candidate with $N$ items $C = \{c_k\}_{1 \leq k \leq N}$, the goal of slate recommendation is to recommend a slate of $K$ items $S = \{i_k\}_{1 \leq k \leq K} \subseteq C$ so as to optimize the overall utility and enhance user experience. We denote the reward of recommending $S$ for user $u$ as $R(S, u)$. Then, the overall slate recommendation problem can be regarded as an combinatorial optimization problem below.

\begin{equation} \label{eq:slate_rec}
\max\limits_S R(S, u;\theta), s.t.  S \subseteq C
\end{equation}
where $\theta$ is the model parameters for generating $S$ from $C$ given user $u$.

If we suppose the user interaction with one item only depends on the quality and relevance of the corresponding item, then we can factorize the slate reward as a weighted sum of the reward for recommending single items and deduce the equation below. 

\begin{equation} \label{eq:slate_factorize}
R(S, u) = \sum\limits_{k=1}^K o_k r(i_k,u), s.t. S = \{i_1,i_2,...,i_K\}
\end{equation}
where $i_k$ the item at position $k$ of the slate, $o_k$ is the observation probability for $i_k$ and $r(i_k,u)$ is the reward for recommending a specific item. As $o_k$ strictly decreases from $1$ to $K$ for most scenarios, the best slate can be constructed by descending sorting the candidate items' reward and returning the top $K$ items. This is exactly the underlying rationality of the existing algorithms in the ranking stage. 

However, given the well-known mutual influence among the slate items \cite{bello2018seq2slate,pang2020setrank,deng2018ad,zhuang2018globally,pei2019personalized,wang2019sequential,jiang2018beyond,zhao2017deep,gong2019exact,wei2020generator,feng2021grn,ai2018learning,ai2019learning,liu2021variation,feng2021revisit}, it is desirable for us to model the cross impact of slate items to optimize the user experience. 

With the multi-stage design of the recommender systems, the matching and pre-ranking stage resort to highly efficient yet coarse-grained models to ensure the coverage rather than precision of the upper stage's candidate set. Moreover, though there are existing solution for the re-ranking stage to consider the mutual influence of slate items, the re-ranking stage has a much smaller candidate set than the ranking stage. Therefore, this work targets to address the modelling of mutual influence among slate items in the ranking stage to further enhance the end to end recommendation experience by improving the quality of the re-ranking stage's candidate set. 
\vspace{-0.2cm}
\subsection{General Design of SAR}

The ranking stage of recommender systems, which normally selects top tens of items from a pool of hundreds to thousands, originally uses model to estimate the reward score for each pair of user-item and forwards the top items to the re-ranking stage. The reward score may depend on a single objective or multiple objectives. For example, the reward may depend on the user click, item watch time, and other satisfaction-related metrics (e.g., liking and sharing) in the content recommendation application. For multi-objective modelling, we usually leverage multi-task learning approaches \cite{tang2020progressive,ma2018modeling} to accurately model multiple user feedback. Furthermore, to compute the overall reward, we need merge the multiple predictions with a function $\Phi$ shown in equation \ref{eq:score_merge} to derive the item's final reward score for greedy ranking.  
\begin{equation} \label{eq:single_obj_pred}
p_1,p_2,...,p_M = f_{base}(i,u)
\end{equation}
\begin{equation} \label{eq:score_merge}
r(i,u) = \Phi(p_1, p_2, ... , p_M)
\end{equation}
where $M$ is the number of prediction tasks, $i$ is the target item for score computation and $f_{base}$ is the DNN model to generate prediction for each task. In addition, $\Phi$ is usually a function manually tuned by the engineers to reflect the reward based on the business goals.

From equation \ref{eq:single_obj_pred}, we can see the original ranking models does not consider the impact of surrounding items on the current one. Ideally, we should estimate $\{p_1,p_2,...,p_M\}$ with the equation \ref{eq:single_obj_pred_ideal}, which takes the recommendation slate $S$ as input. However, we cannot get the full recommendation slate during online serving in the ranking stage.
\begin{equation} \label{eq:single_obj_pred_ideal}
p_1,p_2,...,p_M = f_{ideal}(i,u,S)
\end{equation}

As a result, we propose to estimate the slate information based on the user features. Since the recommender systems tend to recommend a slate of items that suits well for the user's interest, theoretically, we can roughly deduce the slate information with the user features. Specifically, during training, we align the user features and slate features in the transformed latent space, namely $l_u$ for user features and $l_s$ for slate features. During inference, we firstly estimate $l_u$ based on the user features, which is concatenated with the user information and item information to predict $\{p_1,p_2,...,p_M\}$. In this way, the model inference does not rely on the slate features at all. 

\vspace{-0.2cm}
\subsection{Model Architecture}

\subsubsection{Baseline Model}
For the ranking stage, we usually take two types of features as input, namely user and item features. Traditionally, we also need manual cross features so as to describe feature interactions, which are not so prevalent now because the effective implicit cross within the deep learning models \cite{he2017neural,chen2019rafm,xiao2017attentional,he2017neural2,cheng2016wide} can warrant a better balance between model size and recommendation performance. Both of the two types of features are high-dimensional binary features derived from one-hot encoding of categorical variables (e.g., user and item ids) or discretization of dense variables (e.g., user activity counting).  As is displayed in Figure \ref{fig:1}(a), for the baseline model, we firstly project each feature to a dense embedding vector with the fully connected embedding layer. Then, the embeddings are concatenated by group to get $e_i$ for item features and $e_u$ for user features. Then, we concatenate them together and input to the backbone network, which utilizes multi-task algorithms \cite{tang2020progressive,ma2018modeling} for multi-objective predictions or simpler algorithms \cite{he2017neural2,rendle2010factorization,cheng2016wide} for single task prediction. 

The whole process can be formally described with the equations below.

\begin{equation} \label{eq:u_embedding}
e_u = Embedding_u(u)
\end{equation}
\begin{equation} \label{eq:i_embedding}
e_i = Embedding_i(i)
\end{equation}
\begin{equation} \label{eq:orig_backbone}
p_1,p_2,...,p_M = f_{base}(i,u) = Backbone(e_u, e_i)
\end{equation}
where $Embedding_u$ and $Embedding_i$ are the embedding and concatenation operation for the user features and item features respectively. And $Backbone$ denotes the backbone network.

\subsubsection{Introduced Components of SAR}
Please refer to Figure \ref{fig:1}(b) and \ref{fig:1}(c) for the model architecture of SAR. First, with SAR, besides the aforementioned two types of features, we also add slate-wise features, such as the sequence of item ids and categories in the recommendation slate. First, these slate-wise features are projected to low dimensional representation with an embedding layer and then concatenated together to get $e_s$ with equation \ref{eq:s_embedding}.
\begin{equation} \label{eq:s_embedding}
e_s = Embedding_s(S)
\end{equation}

Second, we add two encoder networks, namely $Encoder_u$ and $Encoder_s$, for the user features and slate-wise features respectively to get the transformed latent space to be aligned with. The slate encoder network denoted by Equation \ref{eq:encoder_s} takes $e_s$ and $e_u$ as input and generates the intermediate representation vector ($l_s$) representing the slate information with personalized weight for different slate items. Moreover, the user encoder network denoted by Equation \ref{eq:encoder_u} accepts $e_u$ as input and produces a representation vector ($l_u$) to be aligned with $l_s$ during training.

\begin{equation} \label{eq:encoder_s}
l_s = Encoder_s(e_s, e_u)
\end{equation}
\begin{equation} \label{eq:encoder_u}
l_u = Encoder_u(e_u)
\end{equation}

Additionally, we introduce the decoder network as is described in \ref{eq:decoder}. Besides the auxiliary input of $e_u$, it also accepts $l_u$ as input for inference or $l_s$ as input for training. The decoder network extracts valuable information from the output latent space of the encoder networks for the specific user and tries to enhance the overall recommendation performance.
\begin{equation} \label{eq:decoder}
d = Decoder(l, e_u)
\end{equation}
where $l$ can be $l_u$ or $l_s$ based on the status of inference or training. And $d$ is the decoding result.

Finally, the decoded information ($d$) together with $e_u$ and $e_i$ is then fed to the backbone networks to generate more accurate predictions for the tasks.
\begin{equation} \label{eq:sar_backbone}
p_1,p_2,...,p_M = Backbone(e_u, e_i, d)
\end{equation}

For the slate encoder network ($Encoder_s$), because of the sequential nature of the item permutation in the slate, we may firstly process the sequence of item features with the techniques of sum pooling, target attention \cite{zhou2018deep}, LSTM\cite{greff2016lstm} or transformer \cite{vaswani2017attention}, to extract essential information to be concatenated with $e_u$ for further computation with the MLP network within $Encoder_s$. For all the other newly introduced modules, we just consider simple MLP networks.

\subsubsection{SAR Training}
For \textbf{\textit{SAR}} training displayed in Figure \ref{fig:1}(b), we need all the aforementioned components, including the slate-wise features and the slate encoder networks. The encoding result of the user encoder network and the slate encoder network is aligned with the supervision of the similarity loss. Furthermore, together with the auxiliary input of $e_u$, the decoder network tries to extract valuable information from the output of the slate encoder network for the specific user, which is further concatenated with $[e_i, e_u]$ and fed to the backbone network for prediction. 

\subsubsection{SAR Inference}
For \textbf{\textit{SAR}} inference displayed in Figure \ref{fig:1}(c), we no longer need the slate-wise features and the slate encoder network. The result of the user encoder network rather than the slate encoder network is fed to the decoder network for further processing. The prediction performance is guaranteed with the proper supervision from the similarity loss during training, which enforces little gap and maximum similarity between the outputs of the user encoder network and the slate encoder network.
\vspace{-0.2cm}
\subsection{Joint Loss Optimization}
In multi-task learning, a common formulation of joint loss is the weighted sum of the losses for individual tasks. With our method, we just add an auxiliary similarity loss to ensure the similarity between the outputs of the slate encoder network and user encoder network. Then, for multi-task learning with $M$ tasks, given input features $X$ and task specific labels $Y_m, m=1,2,...,M$, we learn the model parameters by minimizing the aggregated loss of equation \ref{eq:2}. $\omega_m$ and $L_m$ are the weight and loss for the task $m$ respectively. For $Loss_{sim}$, we use the L2 loss to minimize the Euclidean distance of the results of the user encoder network ($l_u$) and the slate encoder network ($l_s$). And $\lambda$ is the loss weight for the similarity loss. 

\begin{equation} \label{eq:2}
Loss(X, Y_{1:M}) = \sum_{m=1}^{M}{\omega_{m}L_{m}(X,Y_m)} + \lambda Loss_{sim}
\end{equation}

It is important for $\lambda$, the weight of the similarity loss, to be tuned and properly set. On the one hand, the model training may not converge with a too large value. Instead, if it is set too small, there will be unfavorable inconsistencies between training and inference. If the model performance is over sensitive to this parameter, there will be difficulties for SAR to be practically applied. We will perform sensitivity test and report the results in our experiments. 
\vspace{-0.2cm}
\section{Experiments}

In this section, we conduct offline and online experiments with the aim of answering the following questions:
\begin{itemize}
\item[\textbf{RQ1}] Compared with the baseline and \textit{PFD} \cite{xu2020privileged}, what is the offline performance of \textbf{\textit{SAR}} on different backbone networks?
\item[\textbf{RQ2}] Can \textbf{\textit{SAR}} achieve additional online gains if deployed on top of the state of the art re-ranking method for slate recommendation?
\item[\textbf{RQ3}] What is the impact of the similarity loss's weight on model performance? How sensitive does the value of $\lambda$ impact the overall recommendation quality?
\item[\textbf{RQ4}] Can the slate encoder and user encoder network generate similar embedding vectors so as to ensure the serving performance when the similarity loss's weight is properly set?
\end{itemize}
\vspace{-0.2cm}
\subsection{Experimental Settings}
\subsubsection{Datasets} 
We experimented\footnote{We open source the code at: https://github.com/BestActionNow/Slate\_Aware\_Ranking} on two real-world datasets. The first one is the public MovieLens-1M\footnote{http://www.grouplens.org/datasets/movielens} dataset consisting of 1 million ratings from 6000 users on 4000 movies. Considering the slate information is absent in MovieLens-1M, we followed the practice of \cite {liu2021variation,jiang2018beyond} to split user rating sessions into slates of size 20 and take the ratings of 4-5 as positive and 1-3 as negative. The data were randomly separated into training set, validation set and test set by the ratio of 8:1:1.

Moreover, we sampled an industrial dataset from 9 days' user logs of our industrial news recommender systems. There are about 30 million users, 2 million items and 1 billion samples in the datasets. In our application, for each user interaction, the system recommends a slate of 10 items, whose ids are used as the slate feature of each exposed item. And our main prediction tasks are the binary classification task of click through rate with cross entropy loss and the regression task of item watch time after click with huber loss. We used the samples in the first 7 days for training. The samples of the 8th day were used for validation and hyper-parameter tuning. And samples of the last remaining day were for testing.


\subsubsection{Experimental Models}

For the slate encoder network at Figure \ref{fig:1}, we experiment several variants of slate feature transformation to study the impact of different slate encoding structures. With these variants, we would like to understand whether the slate's sequential information and the proper attention weights for different slate items are beneficial for the overall recommendation performance. For all of these variants, we concatenate the above result with $e_u$ and input to the MLP component of the slate encoder network. 
\begin{itemize}

\item \textbf{$SAR_{SumPool}$}: We perform sum pooling with the embeddings of the slate items to obtain the slate embedding. For this configuration, neither sequential information nor relative importance of slate items can be captured. 
\item \textbf{$SAR_{LSTM}$}: We input the embeddings of the slate items to a one layer LSTM network to get the slate embedding. For this configuration, the sequential information of slate items is kept. But we do not compute the relative weights of the slate items based on the target item, whose ranking scores are to be predicted.
\item \textbf{$SAR_{Attn}$}: Following \cite{zhou2018deep}, we single out the target item as query to get the attention weight over the slate items. Please note that the target item belongs to the set of slate items. Then, weighted sum is executed to get the slate embedding. In this way, we get the relative weights of the slate items favorable for the target item but lose the slate's sequential information.

\end{itemize}
For the user encoder, slate encoder and decoder network of \textbf{\textit{SAR}}, we use MLP structure of [$dim$, $dim$] where $dim$ is set to 16 and 128 for MovieLens-1M and the industrial data respectively. For the backbone network at Figure \ref{fig:1}, PLE \cite{tang2020progressive} is utilized for the industrial dataset. And multiple backbone networks, including FM\cite{rendle2010factorization}, Wide\&Deep\cite{cheng2016wide} and NCF\cite{he2017neural2}, are tested for MovieLens-1M to verify the wide compatibility of our method.

And we also take the slate-wise features as privileged features, which are discriminative but only available during training, and test the performance of privileged feature distillation (\textbf{\textit{$PFD$}}) \cite{xu2020privileged}. For fairness, we would like the teacher model of PFD to have the same capacity as the model for SAR training. Thus, after disabling the similarity loss, we use the model structure of Figure \ref{fig:1}(b) with the $SAR_{Attn}$ slate encoder network as teacher model and the model structure of Figure \ref{fig:1}(a) as student model.

For all the models, we use Adam \cite{kingma2014adam} optimizer. We tune the learning rate and other hyper-parameters for each model separately.

\begin{figure}
\setlength{\abovecaptionskip}{0.0cm} 
	\centering\textbf{}
	\includegraphics[scale=0.35]{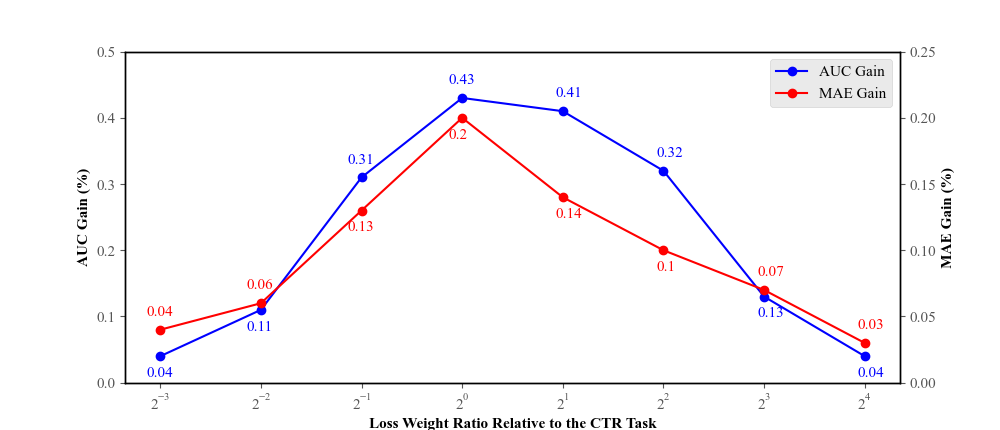}
	\caption{The impact of similarity loss weight}
	\label{fig:2}
\end{figure}

\subsubsection{Evaluation Metrics}
From equation \ref{eq:sar_backbone} and \ref{eq:orig_backbone}, we can see that the main task of the ranking model is to predict one or several user feedback, such as clicking, watch time, liking and sharing. And the merge function $\Phi$ in Equation \ref{eq:score_merge}, which is a function manually tuned to reflect the reward based on the business goals, assumes that the ranking model can estimate accurate interaction probabilities for binary classification tasks (e.g. liking and sharing) and absolute values for regression tasks (e.g. watch time). 

As a result, instead of the ranking metrics, such as Normalized Discounted Cumulative Gain (NDCG) \cite{wang2013theoretical} and Mean Reciprocal Rank (MRR) \cite{enwiki:1095286224}, we use the metrics of Area Under the ROC Curve (AUC) \cite{flach2011coherent} for classification tasks and Mean Absolute Error (MAE) \cite{enwiki:1087554218} for regression tasks. Please note that many other recommendation literature, such as \cite{tang2020progressive}, also use similar metrics. For AUC, a bigger value indicates better performance. While, for MAE, it is the smaller the better.

\begin{figure*} 
  \centering 
  \subfigure[general distribution]{ 
    \label{fig:3:a} 
    \centering
    \includegraphics[width=2.5in]{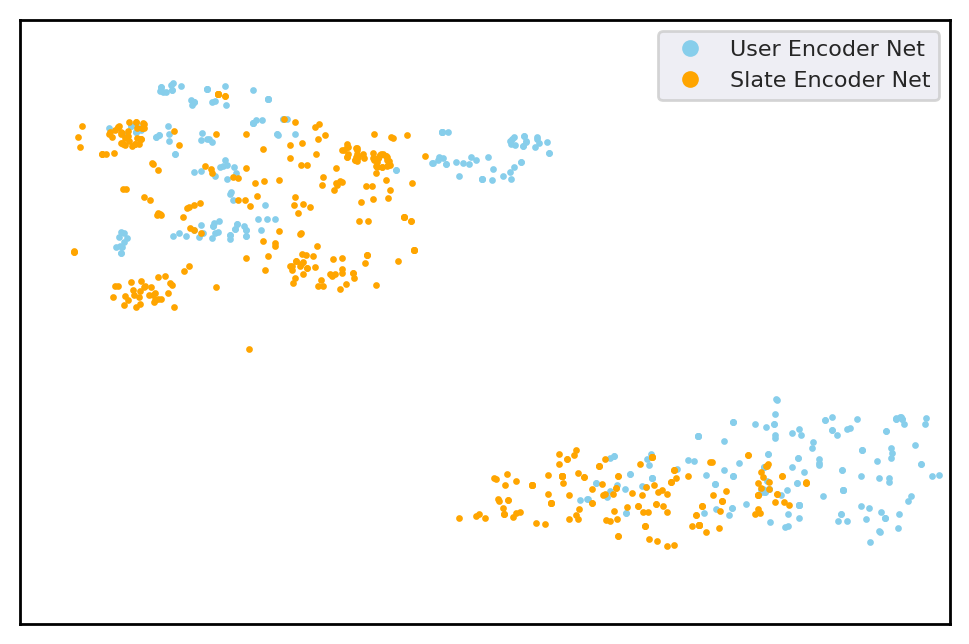} 
    }
    \subfigure[sampled pairs]{ 
    \label{fig:3:b} 
    \centering
    \includegraphics[width=2.5in]{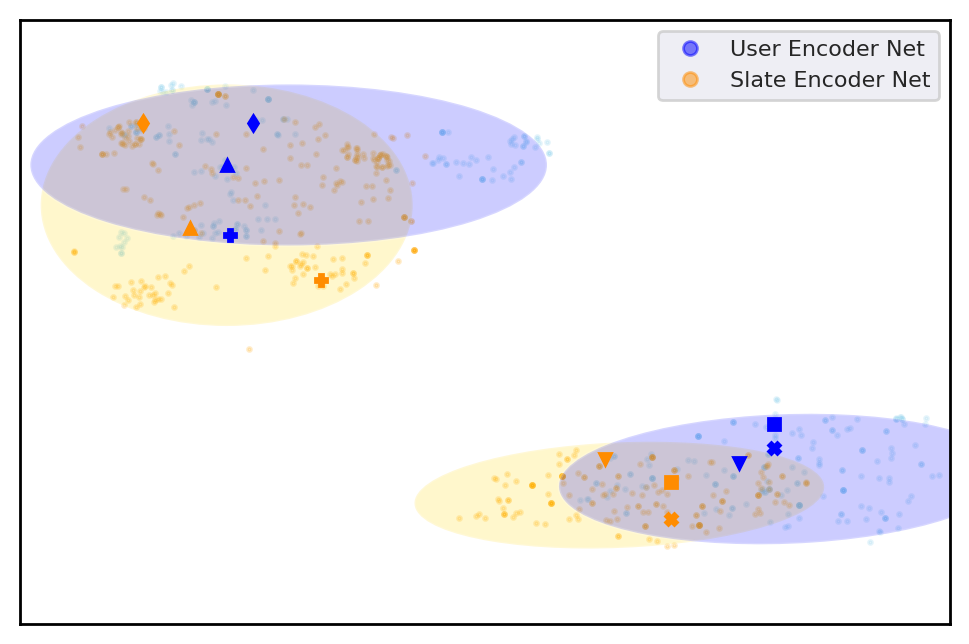} 
    }
    \subfigure[general distribution 1st dim]{ 
    \label{fig:3:c} 
    \centering
    \includegraphics[width=2.5in]{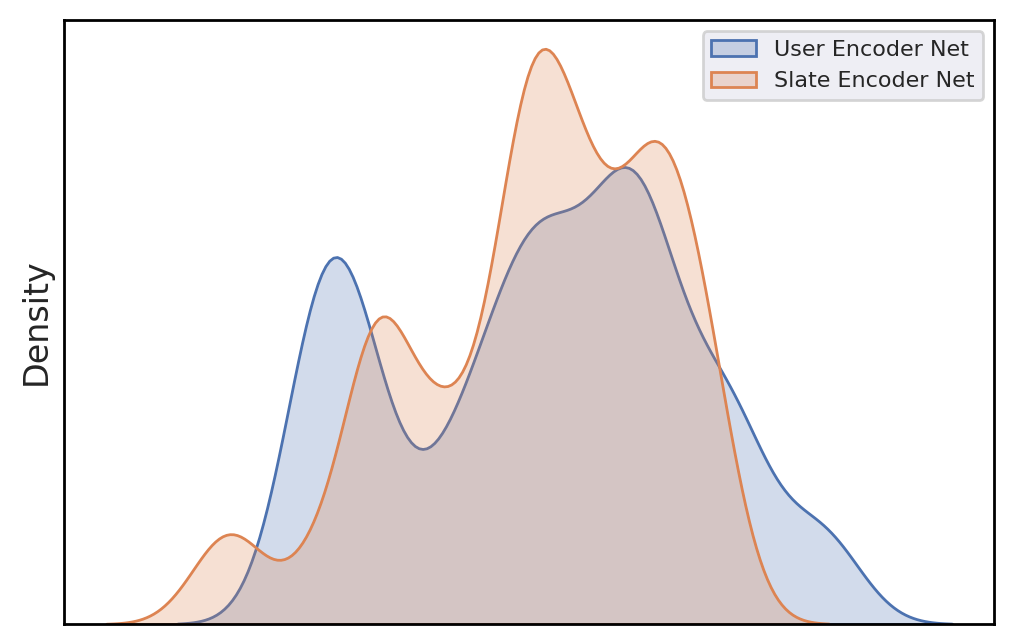} 
    }    
    \subfigure[general distribution 2nd dim]{ 
    \label{fig:3:d} 
    \centering
    \includegraphics[width=2.5in]{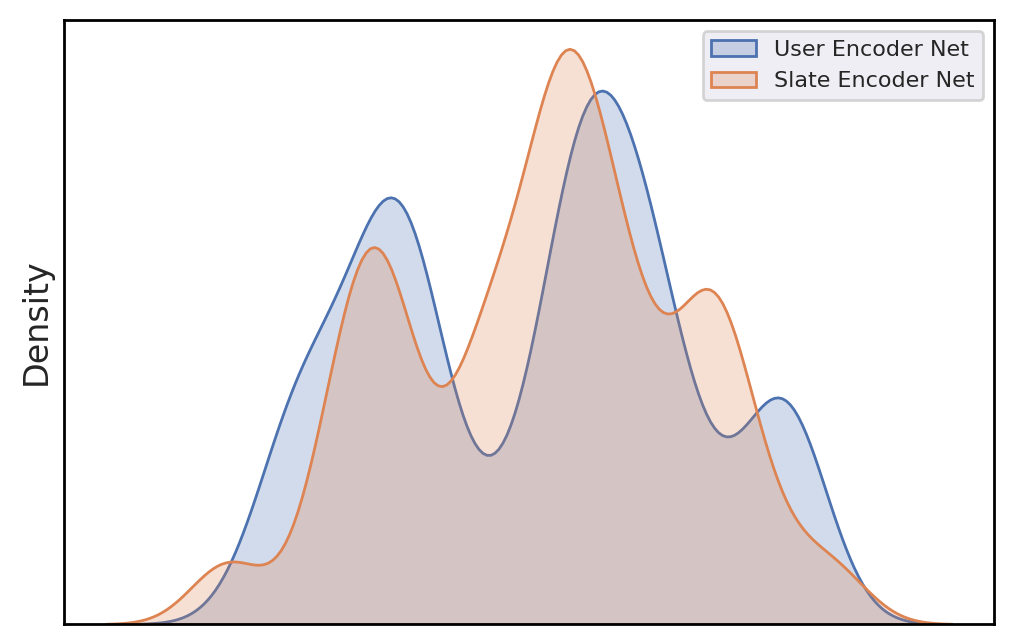} 
    } 
    \caption{Encoder Output Visualization}
    \label{fig:3}
\end{figure*}

\vspace{-0.2cm}
\subsection{Offline Performance Comparison(\textbf{RQ1})}

As mentioned above, we use AUC and MAE as the evaluation metrics for the tasks of classification and regression respectively. Please note that a 0.1$\%$ relative gain over baseline means significant performance improvement for both metrics in our scenario. 

\begin{table}[H]
\centering
\caption{Offline Performance for Industrial Dataset}
\label{tab:offline}
\begin{tabular}{ |l|cc|cc| } 
\hline
$Model$ & $AUC_{ctr}$ & $RelGain$ & $MAE_{wt}$  & $RelGain$ \\
\hline
$Baseline$ &  0.7841 & 0$\%$ & 11.3858  & 0$\%$\\
$PFD_{student}$\cite{xu2020privileged}  & 0.7849 & +0.10$\%$ & 11.3760  & +0.09$\%$ \\
$SAR_{SumPool}$  & 0.7871 & +0.40$\%$ & 11.3672  & +0.16$\%$ \\
$SAR_{LSTM}$ & 0.7872 & +0.41$\%$ & 11.3720   & +0.12$\%$ \\
$SAR_{Attn}$ & \textbf{0.7875} & \textbf{+0.43$\%$} & \textbf{11.3631}  & \textbf{+0.20$\%$} \\
\hline
\end{tabular}
\end{table}
\subsubsection{Industrial Dataset}

The baseline's architecture is shown in Figure \ref{fig:1}(a) with PLE \cite{tang2020progressive} as its backbone network. For PFD, the student model is of the same structure as the baseline. But its student model is trained with the supervision from both the soft labels of the teacher model and the original hard labels. For the teacher model of PFD, after removing the similarity loss, we use the model structure of Figure \ref{fig:1}(b) with $SAR_{Attn}$ as the slate encoder network. For SAR, we use different model structures for training and inference, which is described in section 3. For the backbone network of SAR, it is of the same structure as the backbone network of the baseline except for the input layer, which is of additional dimension with the introduction of $d$ in Equation \ref{eq:sar_backbone}. 

From Table~\ref{tab:offline}, we can see that our method achieves competitive performance. Compared with baseline, $SAR_{Attn}$ has 0.43$\%$ gain on AUC and 0.2$\%$ decrease on MAE. $SAR_{Attn}$ performs better than $PFD_{student}$ with 0.33$\%$ gain on AUC and 0.11$\%$ decrease on MAE. In other words, $SAR$ significantly outperforms $PFD$ with about half training resources. 

Among the variants of $SAR$, $SAR_{Attn}$ shows some marginal gains over $SAR_{LSTM}$ and $SAR_{SumPool}$. Based on our test, the gain is stable for multiple runs. Moreover, $SAR_{LSTM}$ and $SAR_{SumPool}$ have roughly the same performance. Therefore, it seems the computed relative weights of the slate items are useful for model prediction. 
\vspace{-0.2cm}
\subsubsection{MovieLens-1M}
For MovieLens-1M, we compare $SAR_{Attn}$, which exhibits the best performance for the industrial data,  with baseline and $PFD$ on three backbone networks. For PFD, the student model is still of the same structure as the baseline model shown in Figure \ref{fig:1}(a). And the teacher model of PFD uses the model structure of Figure \ref{fig:1}(b) with $SAR_{Attn}$ as the slate encoder network after disabling the similarity loss. For SAR, we still use different model structures for training and inference shown at Figure \ref{fig:1}(b) and Figure \ref{fig:1}(c) respectively. For this experiment, we test three different backbone networks listed in table \ref{tab:ml} to study SAR's performance gain over different backbone networks.

The results are summarized in Table~\ref{tab:ml}. Compared with the baseline, PFD shows gain for the backbone network of Wide\&Deep and NCF but neural results for FM. Moreover, SAR shows consistent gains over PFD and the baseline model. Overall, the results of Movie-Lens 1M demonstrate the effectiveness and wide compatibility of $SAR$. 
\begin{table}
\centering
\caption{Offline Performance for MovieLens-1M}
\label{tab:ml}
\begin{tabular}{|l|c|cc|cc|} 
\hline
\multirow{2}{*}{Backbone} & $Baseline$ & \multicolumn{2}{c|}{$PFD_{student}$} & \multicolumn{2}{c|}{$SAR_{Attn}$}  \\ 
\cline{2-6}
                          & $AUC$      & $AUC$  & $RelGain$                   & $AUC$  & $RelGain$                 \\ 
\hline
FM\cite{rendle2010factorization}                        & 0.8065     & 0.8062 & -0.03\%                     & \textbf{0.8079} & \textbf{+0.17\%}                   \\
Wide\&Deep\cite{cheng2016wide}                  & 0.8055     & 0.8063 & +0.10\%                     & \textbf{0.8087} & \textbf{+0.40\%}                   \\
NCF\cite{he2017neural2}                    & 0.8062     & 0.8086 & +0.30\%                     & \textbf{0.8110} & \textbf{+0.60\%}                   \\
\hline
\end{tabular}
\end{table}


\vspace{-0.2cm}
\subsection{Online A/B Testing(\textbf{RQ2})}

To verify the effectiveness of our proposed algorithm, we served $SAR_{Attn}$ model online in the ranking stage, which accepts hundreds of candidates and outputs the top 50 items to the re-ranking stage, of our content recommender systems. In the re-ranking stage, we already deployed the state of the art \textbf{\textit{generator and evaluator}} algorithms similar to \cite{feng2021revisit} for slate recommendation. We randomly distributed online users to two buckets with the baseline ranking model or \textbf{\textit{SAR}} and evaluated the performance for seven days to report our main online metrics of total page view per person (PV) and app stay time per person (ST). As is shown in table \ref{tab:online}, \textbf{\textit{SAR}} achieved significant gains over the baseline by \textbf{0.9$\%$} for page view and \textbf{0.745$\%$} for stay time. Based on the results from our A/B test platform, the p-value is below 0.05 for one-tailed t-test for both metrics mentioned above.

Unlike search engine, the user's intention is not so focused in recommendation scenario. Thus, after implicitly modelling the mutual influence of the slate items with \textbf{\textit{SAR}}, the recommendation results should cover more interests of the user in a recommendation slate guided by overall utility optimization, thereby exhibiting enhanced diversity. We are concerned about diversity because of the following reasons. On the one hand, diverse recommendation is helpful for immediate user satisfaction. On the other hand, in all likelihood, the system can understand more multi-faceted user interests so as to ensure even better recommendation performance in the future. Furthermore, diverse recommendation will encourage the authors to provide a wide variety of contents, which is beneficial for the long term success of the ecosystem. To verify SAR's effectiveness on diversity, we extracted one day's online data and analyzed the final recommendation results. We computed the exposed items' gini index \cite{enwiki:1101796604} metrics on item id and item category. From table \ref{tab:gini}, compared with the baseline traffic, we can see lower gini index indicating better diversity for the final recommendation results, especially for the metrics of item category.

\begin{table}[H]
\centering
\caption{Online Performance}
\label{tab:online}
\begin{tabular}{ |l|c|c| } 
\hline
 & App Stay Time per Person & Page View per Person \\
\hline
$Gain$ & \textbf{0.745}$\%$ & \textbf{0.9}$\%$ \\
\hline
\end{tabular}
\end{table}
\begin{table}[H]
\centering
\caption{The Gini Index of Online Traffic}
\label{tab:gini}
\begin{tabular}{ |l|cc|cc| } 
\hline
$Model$ & $ItemID$ & $RelGain$ & $ItemCat$ & $RelGain$ \\
\hline
$Baseline$ & 0.8829 & 0$\%$ & 0.7253 & 0$\%$ \\
$SAR_{Attn}$ & \textbf{0.8818} & \textbf{+0.12}$\%$ & \textbf{0.7126} & \textbf{+1.75}$\%$  \\
\hline
\end{tabular}
\end{table}

\vspace{-0.2cm}
\subsection{Impact of Similarity Loss Weight(\textbf{RQ3})}

We further studied the impact of the weight of similarity loss on model performance in the industrial dataset. First, following the results of previous experiments, we fixed the loss weights of the CTR task and watch time task to the optimal values. Then, to report the model performance of $SAR_{Attn}$, we tested different weight ratio of the similarity loss relative to the CTR task. 

According to Figure \ref{fig:2}, the best model performance is achieved when the similarity loss's weight is equal to the CTR task. Moreover, $SAR_{Attn}$'s performance is not too sensitive to the change of the similarity loss weight. So hyper-parameter tuning is of manageable difficulty and \textit{SAR} can be practically applied to boost the recommendation performance. 

\vspace{-0.2cm}
\subsection{Encoder Network Embedding Distribution Analysis on Industrial Dataset(\textbf{RQ4})}

We visualized the embedding distribution of the User and Slate Encoder Networks with t-SNE \cite{van2008visualizing} in figure \ref{fig:3:a}, \ref{fig:3:c} and \ref{fig:3:d} for the best performing $SAR_{Attn}$ model with equal loss weights for the similarity task and the CTR task. From figure \ref{fig:3:a}, we can see the points fall into two clusters and the embeddings from the two encoders show similar distribution. With figure \ref{fig:3:c} and \ref{fig:3:d}, it is obvious that the distribution density match well in each dimension.

In figure \ref{fig:3:b}, we randomly sampled six output pairs from these two encoder networks. For each pair, we got the encoding results from the two encoder networks for the same instance. And we plotted with different colors for points from different networks. The points from the same pair were of the same shape. For each pair, the two points resided in the same cluster and were close to each other.

In summary, the User Encoder network and Slate Encoder Network can generate outputs with consistent distributions, which ensures the consistency between training and inference and explains the effectiveness of our proposed method.


\vspace{-0.2cm}
\section{Conclusion}
In this paper, we propose an algorithm for the ranking stage of recommender systems to predict user behaviors more accurately by implicitly modelling the contextual items' impact. We perform thorough offline experiments and online A/B testing to prove that it significantly enhances the recommendation relevance and diversity by ensuring the quality of the re-ranking stage's candidates. Moreover, additional experiments are carried out to study the similarity loss weight's impact and understand why $SAR$ works.  In future, we will try to further improve the recommendation performance by designing more effective models in the ranking and re-ranking stage to tackle the issue of cross item influence. Furthermore, though the matching and pre-ranking stage care more about coverage than precision, it is interesting to investigate whether addressing the aforementioned issue in these two layers can boost the overall recommendation quality. Finally, as it is essential to properly handle position bias for precise user preference modelling \cite{ren2022unbiased,joachims2017unbiased,ai2018unbiased}, we plan to study how to more accurately estimate the context items' impact on target items at different positions.


\bibliographystyle{ACM-Reference-Format}
\balance
\bibliography{references}

\end{document}